# Reproducibility Made Easy: A Tool for Methodological Transparency and Efficient Standardized Reporting based on the proposed MRSinMRS Consensus.

**Antonia Susnjar[1*], Antonia Kaiser[2*], Dunja Simicic[3,4], Gianna Nossa[2], Alexander Lin[5], Georg Oeltzschner[3,4], Aaron T. Gudmundson[3,4,6]**

**1)** Athinoula A. Martinos Center for Biomedical Imaging, Institute for Innovation in Imaging, Department of Radiology Massachusetts General Hospital and Harvard Medical School, Boston MA.
**2)** CIBM Center for Biomedical Imaging, École polytechnique fédérale de Lausanne (EPFL), Lausanne, Switzerland
**3)** Russell H. Morgan Department of Radiology and Radiological Science, The Johns Hopkins University School of Medicine, Baltimore, MD.
**4)** F. M. Kirby Research Center for Functional Brain Imaging, Kennedy Krieger Institute, Baltimore, MD.
**5)** Center for Clinical Spectroscopy, Department of Radiology, Brigham and Women's Hospital, Harvard Medical School, Boston, MA.
**6)** The Malone Center for Engineering in Healthcare, Johns Hopkins University, Baltimore MD.

* authors have contributed equally

Corresponding Author:
Aaron Gudmundson
agudmun2@jhmi.edu
Johns Hopkins University School of Medicine
733 N Broadway, Baltimore, MD 21205

**Abstract**

Recent expert consensus publications have highlighted the issue of poor reproducibility in magnetic resonance spectroscopy (MRS) studies, mainly due to the lack of standardized reporting criteria, which affects their clinical applicability. To combat this, guidelines for minimum reporting standards (MRSinMRS) were introduced to aid journal editors and reviewers in ensuring the comprehensive documentation of essential MRS study parameters. Despite these efforts, the implementation of MRSinMRS standards has been slow, attributed to the diverse nomenclature used by different vendors, the variety of raw MRS data formats, and the absence of appropriate software tools for identifying and reporting necessary parameters. To overcome this obstacle, we have developed the REproducibility Made Easy (REMY) standalone toolbox. REMY software supports a range of MRS data formats from major vendors like GE (p. file), Siemens (Twix, .rda, .dcm), Philips (.spar/.sdat), and Bruker (.method), and MRS-NIfTI (.nii/.json) files (.nii/.nii.gz) facilitating easy data import and export through a user-friendly interface. REMY employs external libraries such as spec2nii and pymapVBVD to accurately read and process these diverse data formats, translating complex header information into a comprehensive structure that adheres to consensus reporting standards, thereby ensuring compatibility and ease of use for researchers in generating reproducible MRS research outputs. Users can select and import datasets, choose the appropriate vendor and data format, and then generate an MRSinMRS table, log file, and methodological documents in both Latex and PDF formats. No coding knowledge is required, making the tool accessible to a wider range of users, including researchers and clinicians without programming expertise. This eliminates technical challenges related to data formatting and reporting. REMY effectively populated key sections of the MRSinMRS table with data from all supported file types. Accurate generation of hardware parameters including field strength, manufacturer, and scanner software version were demonstrated. However, it could not input data for RF coil and additional hardware information due to their absence in the files. For the acquisition section, REMY accurately read and populated fields for the pulse sequence name, nominal voxel size, repetition time (TR), echo time (TE), number of acquisitions/excitations/shots, spectral width [Hz], and number of spectral points, significantly contributing to the completion of the 'Acquisition' fields of the table. Furthermore, REMY generates a boilerplate methods text section for manuscripts.The use of REMY will facilitate more widespread adoption of the MRSinMRS checklist within the MRS community, making it easier to write and report acquisition parameters effectively.

**Acronyms**:

aux – Auxiliary (LaTeX Document)
bbl – Bibliography (LaTeX Document)
bib – Citations (LaTeX Document)
bcf – Control file (LaTeX Document)
blg – Bibliography log (LaTeX Document)
CSV – Comma separated values
GUI – Graphical user interface
HERMES – Hadamard Encoding and Reconstruction of MEGA-Edited Spectroscopy
JSON – JavaScript Object Notation
LASER – Localized adiabatic selective refocusing
MEGA – MEscher GArwood
MR – Magnetic Resonance
MRS – Magnetic Resonance Spectroscopy
MRSinMRS – Minimum Reporting Standards for in vivo Magnetic Resonance Spectroscopy
REMY – Reproducibility Made Easy

SVS – Single Voxel Spectroscopy
TR – Repetition time
TE – Echo time
NA – Number of acquisitions or excitations per spectrum
OS – Operating system
VOI – Volume of interest
PDF – Portable Document Format
PRESS – Point RESolved Spectroscopy
sLASER – Semi-adiabatic localized adiabatic selective refocusing
STEAM – Stimulated Echo Acquisition Mode

**Data Availability Statement**
Data sharing is not applicable to this article as no new data were created or analyzed in this study.

## 1. Introduction

Magnetic resonance spectroscopy (MRS) is a non-invasive magnetic resonance (MR) method used to evaluate the metabolic profiles of tissues.[1,2] Over the past four decades, the field of MRS progressed greatly in protocol development for metabolite acquisition, as well as advancements in pre- and post-processing of data and metabolite quantification.[2–6] These efforts have led to a plethora of MRS consensus papers that provide a platform for education and training, standardization, and addressing conflicting viewpoints (Table 1). The combined progress in technical innovations within the MRS field and the consensus achieved among experts aimed to promote MRS to clinical researchers and psychiatrists and to improve access to non-expert users.[3,4,7–18] One particularly impactful consensus publication addressed the urgent need for standardized reporting criteria for MRS studies. The authors recognized that the enormous methodological heterogeneity in the field has contributed to a lack of reproducibility across different groups, sites, and techniques. Furthermore, they identified a lack of structured methods reporting, making it difficult to compare and integrate quantitative results. In fact, the majority of journals within the MR field require study overview checklists for experimental setup details and participant characteristics, such as STARD[19], CONSORT[20], PRISMA[21], and STROBE[22], however, these do not cover reporting standards for acquisition parameters and the technical details that are required for an MR acquisition, including acquisition parameters necessary for MRS measurements. Consequently, comparing studies is difficult and may lead to inaccurate and inconsistent conclusions, contributing to the slow translation of MRS from the research setting to clinical applications.

| Author | Title | Cited |
|---|---|---|
| Oz et al. (2014) | Clinical proton MR spectroscopy in central nervous system disorders | 613 |
| Mullins et al. (2014) | Current practice in the use of MEGA-PRESS spectroscopy for the detection of GABA | 542 |
| Wilson et al. (2019) | Methodological consensus on clinical proton MRS of the brain: Review and recommendations | 274 |
| Near et al. (2021) | Preprocessing, analysis and quantification in single-voxel magnetic resonance spectroscopy: experts' consensus recommendations | 183 |
| **Lin et al. (2021)** | **Minimum reporting standards for in vivo magnetic resonance spectroscopy (MRSinMRS): experts' consensus recommendations** | **104** |
| Cudalbu et al. (2021) | Contribution of macromolecules to brain 1H MR spectra: Experts' consensus recommendations | 83 |
| Choi et al. (2021) | Spectral editing in 1H magnetic resonance spectroscopy: Experts' consensus recommendations | 71 |
| Maudsley et al. (2021) | Advanced magnetic resonance spectroscopic neuroimaging: Experts' consensus recommendations | 71 |
| Meyerspeer et al. (2021) | 31 P magnetic resonance spectroscopy in skeletal muscle: Experts' consensus recommendations | 67 |
| Kreis et al. (2020) | Terminology and concepts for the characterization of in vivo MR spectroscopy methods and MR spectra: Background and experts' consensus recommendations | 64 |
| Juchem et al. (2020) | B0 shimming for in vivo magnetic resonance spectroscopy | 57 |
| Krššák et al. (2020) | Proton magnetic resonance spectroscopy in skeletal muscle: Experts' consensus recommendations | 40 |
| Andronesi et al. (2021) | Motion correction methods for MRS: experts' consensus recommendations | 36 |
| Tkac et al. (2021) | Water and lipid suppression techniques for advanced 1H MRS and MRSI of the human brain: Experts' consensus recommendations | 35 |
| Lanz et al. (2021) | Magnetic resonance spectroscopy in the rodent brain: experts' consensus recommendations | 12 |

Table 1. Compilation of MRS consensus papers, along with their corresponding citation numbers. This table is a resource for any questions within the MRS field and the most updated list on published consensus in MRS.

To overcome the lack of information in publications, a checklist of minimum requirements for reporting MRS studies was developed - the Minimum Reporting Standards for in vivo Magnetic Resonance Spectroscopy (MRSinMRS).[15] The objective was that, for clinical and research studies using MRS, authors should complete the provided table and attach it as a supplementary figure or appendix to their manuscript. This was intended to foster standardization and simplification of the review process and ensure that all critical parameters of data acquisition and analysis are appropriately reported. As a further step, the editors of 121 clinical journals were asked to consider implementing the MRSinMRS table as a requirement for publication, of which seven agreed and 12 are currently evaluating the proposal.[23]

While the expert consensus papers have been widely cited in MRS publications (Table 1), the actual recommendations are not nearly as well implemented in practice. Research groups with long-established workflows experience the adoption of consensus practices as an inconvenience, often due to a lack of strong incentive, while researchers new to the MRS field may not yet have the necessary expertise or guidance to implement them. For example, the MRSinMRS consensus paper, published in 2021, has been cited 102 times to date, but only 43 references actually incorporated the MRSinMRS table, while the remaining 59 citations only acknowledged the paper (Figure 1). While the MRSinMRS table offers a practical solution for reporting methods, locating required parameters within DICOM headers or MRS raw files can be challenging, especially for beginners. This is primarily due to variations in MRS data formats and nomenclature across different vendors and the complex nature of the parameters themselves.[9] Although MRS software developers have begun implementing the MRSinMRS checklist into their standard output tables, community members expressed a clear need for a user-friendly and easy-to-use standalone solution at the 2022 Magnetic Resonance Spectroscopy workshop in Lausanne, Switzerland.[24] Incorporating the MRSinMRS table provides editors, reviewers, and readers with a clear overview of the specific MRS methodology used in each study, while also guaranteeing the availability of comprehensive details for those seeking to replicate an experiment or conduct meta-analyses based on the results. Moreover, the checklist (Figure 2) will establish a consistent format for presenting MRS information and offer journals less acquainted with MRS a coherent means of validating methods.

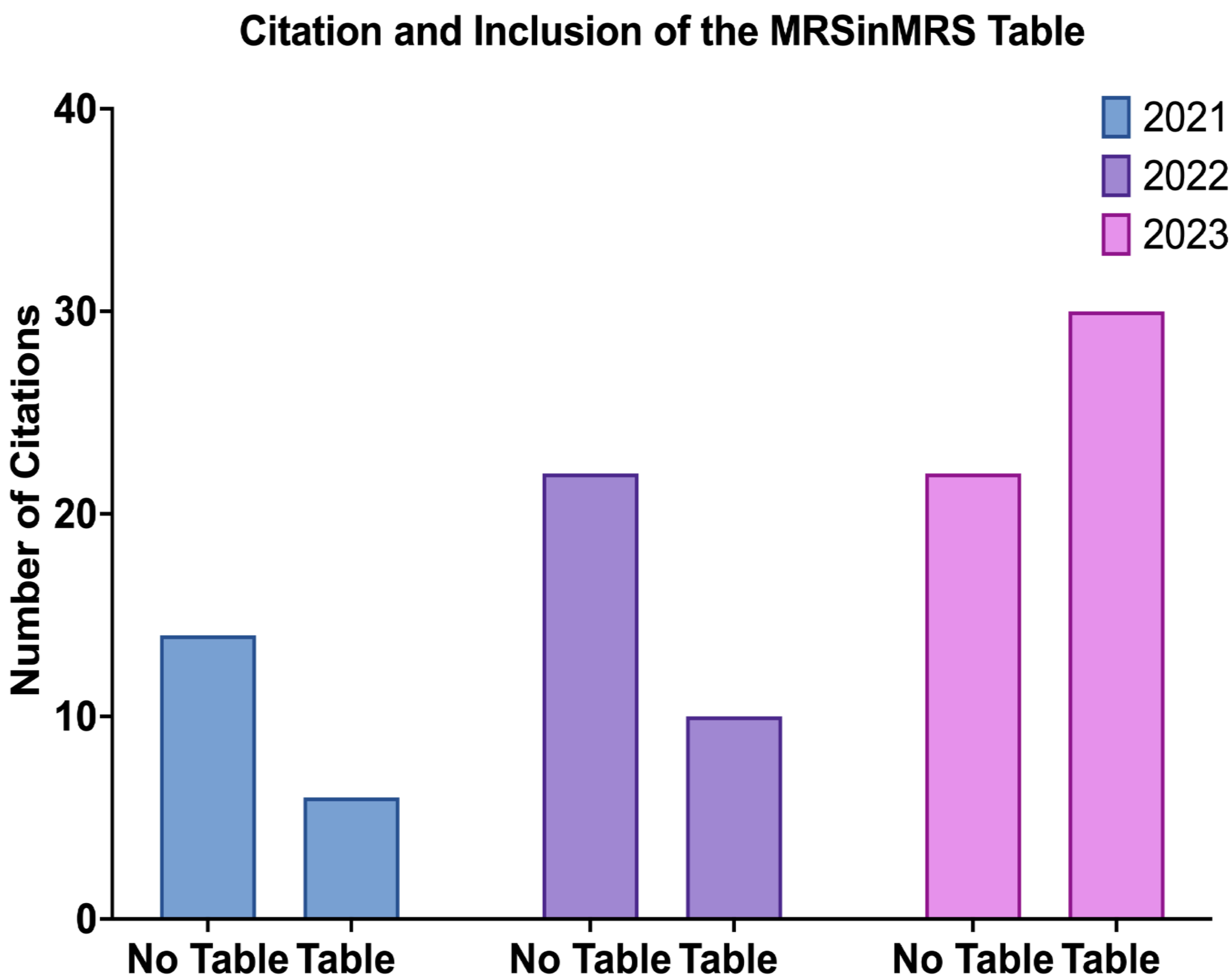

Figure 1. Adoption and Utilization of the MRSinMRS Consensus in MRS Research. This figure illustrates the citation dynamics of the MRSinMRS consensus since its publication in 2021, highlighting a total of 102 citations. It differentiates between papers that merely cited the consensus (n=59) and those that actively integrated its reporting table into their methodology (n=43). The trend indicates an initial phase of citations without substantial adoption of the reporting table in 2021 and 2022. In contrast, 2023 marks a significant shift towards implementing the consensus framework, with 30 papers including the reporting table versus 20 that only cited the consensus. This trend signifies a growing commitment within the MRS research community towards enhancing research reproducibility through standardized reporting. REMY is anticipated to further streamline the reporting process, enabling researchers to automatically populate the table, thereby facilitating more efficient and accurate adherence to the consensus guidelines.

| **MRSinMRS checklist**<br>**Site (name or number)** | ■ Filled by REMY<br>■ Not Available |
|---|---|
| **1. Hardware** | |
| a. Field strength [T] | |
| b. Manufacturer | |
| c. Model (software version if available) | |
| d. RF coils; nuclei (transmit/receive), number of channels, type, body part | |
| e. Additional hardware | |
| **2. Acquisition** | |
| a. Pulse sequence | |
| b. Volume of interest (VOI) locations | |
| c. Nominal VOI size [$cm^3$, $mm^3$] | |
| d. Repetition time ($T_R$), echo time ($T_E$) [ms, s] | |
| e. Total number of excitations or acquisitions per spectrum | |
| In time series for kinetic studies | |
| i. Number of averaged spectra (NA) per time point | |
| ii. Averaging method (eg block-wise or moving average) | |
| iii. Total number of spectra (acquired/in time series) | |
| f. Additional Sequence Parameters (spec. width in Hz, number of spec. points, freq. offsets) | |
| if STEAM:; mixing time ($T_M$) | |
| if MRSI: 2D or 3D, FOV in all directions, matrix size, acc. factors, sampling method | |
| g. Water suppression method | |
| h. Shimming method, reference peak, and thresholds for "acceptance of shim" chosen | |
| i. Triggering or motion corr. method (resp., periph., cardiac trig., incl. device and delays) | |

Figure 2. Abbreviated MRSinMRS table[15] providing an overview of required "Hardware" and "Acquisition" parameters for the publication of MRS studies which are extracted by REMY from vendor raw or deidentified NIfTI data. Blue fill are the fields that REMY populates when the information is present within the data header. Grey fill are fields that are not available in the header or currently under development (i.e., MRSI). A template Excel spreadsheet of this table can be found at: https://github.com/agudmundson/mrs_in_mrs.

To address this need, we have developed an open-source standalone software application to enhance accessibility and streamline the reporting process for both novice and expert MRS researchers. This application automates the population of the hardware and acquisition portions of the MRSinMRS table using a single data file from an MRS study and generates a corresponding methods section to be used in publications. Advantages of the application are evident in easy data input, using an intuitive graphical user interface (GUI) as shown in Figure 3, and freedom from any proprietary dependencies. Furthermore, it offers a solid foundation that the community can build upon to tackle new challenges within the MRS field.

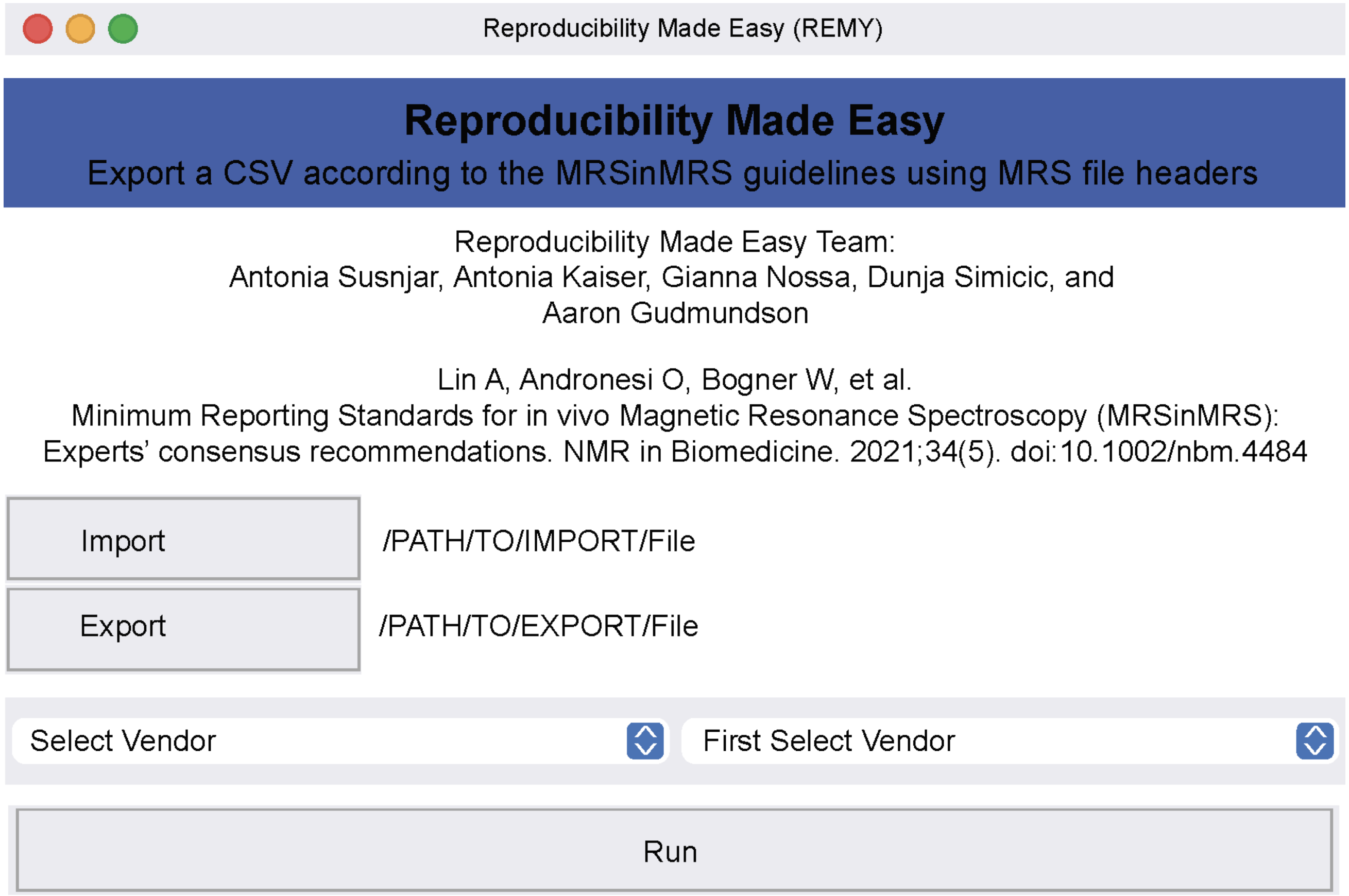

Figure 3. Reproducibility Made Easy User Interface. Single file import facilitates the table and methods section output at the location specified by the user.

## 2. Methods

Here, we describe the development of a software suite termed "REproducibility Made EasY" (REMY). REMY is freely available for download from the GitHub repository https://github.com/agudmundson/mrs_in_mrs; the source code is open under a liberal BSD-3 license. REMY is designed as a standalone application to create the MRSinMRS table and a matching MRS methods section that can be used in publications. The Python-based (v. 3.11) application requires no programming experience, operating through an intuitive graphical user interface (GUI) built using Tkinter[25] (Figure 4). While REMY is available to run through the command line, executables were created using PyInstaller for Windows, macOS, and Linux. The application is operating-system (OS) agnostic, meaning it operates uniformly across platforms. As an open-source application, REMY is transparent, granting visibility into its underlying codebase.

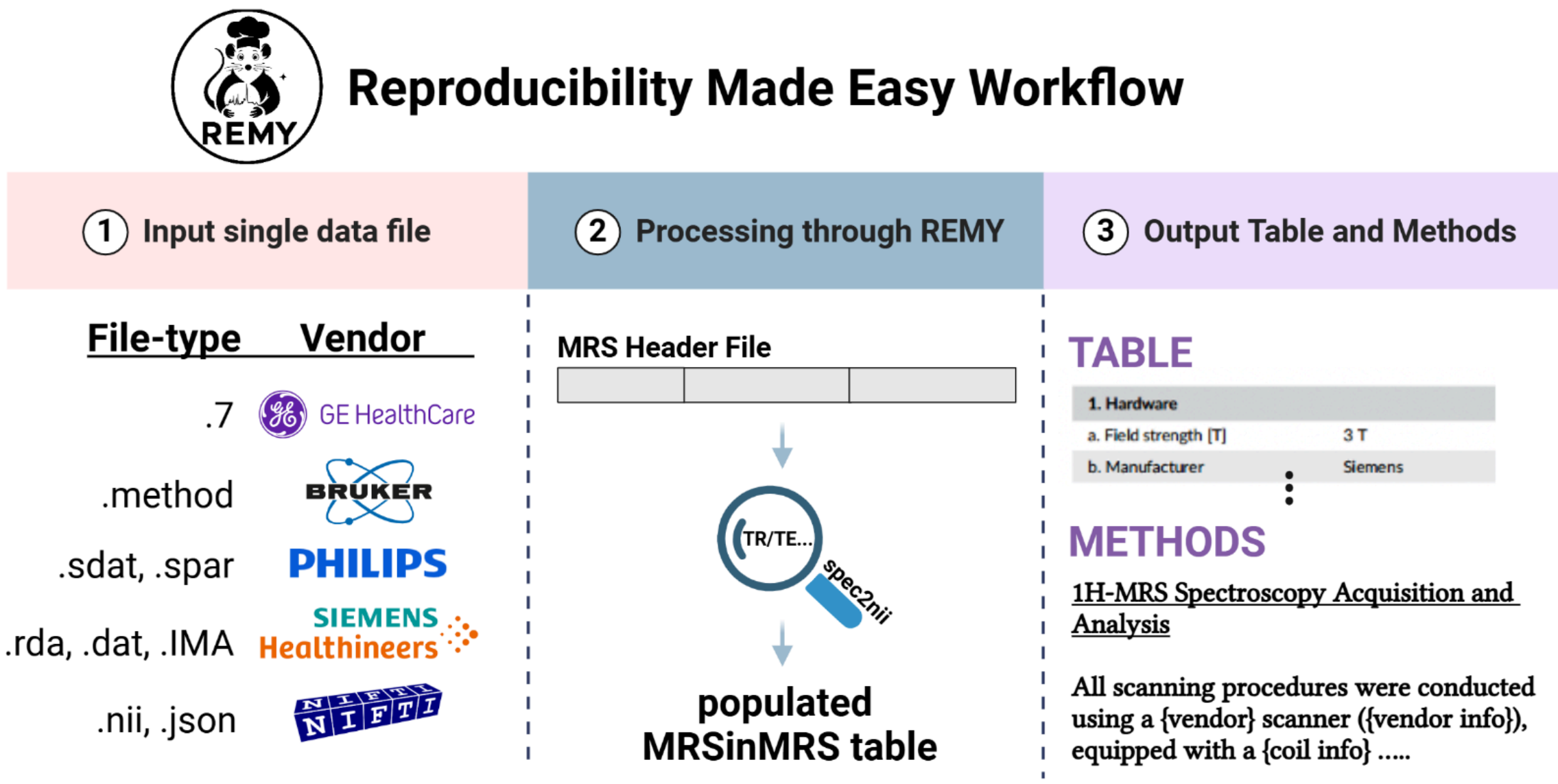

Figure 4. REMY Workflow. REMY supports multiple MRS data formats as input, including GE (.7), Siemens (.ima, .rda, .dat), Philips (.spar/.sdat), Bruker (.method), and NIFTI (.nii). Users initiate the process by importing a dataset, after which REMY looks for the necessary parameters in the header file of the data file, using spec2nii functionality. It then populates the hardware and acquisition parts of the MRSinMRS table and creates documentation (in Latex and PDF formats) for the Methods section.

### Workflow Overview

REMY currently supports various commonly used MRS data formats, including GE pfiles (.7); Siemens DICOM (.ima), .rda, and Twix (.dat); Philips spar/sdat (.spar/.sdat); Bruker Method (.method); and MRS-NIfTI files (.nii/.json). To begin, users will select a dataset with a file explorer using the 'Import' button. Once selected, REMY will automatically update the outputs to export to this directory. Users can also select a different folder using the 'Export' button. Now that a dataset has been selected, users must select which vendor and which data format they have input using the dropdown menus. Finally, the 'Run' button exports the MRSinMRS table, log file, and text documents (LaTeX and PDF format) with a completed and referenced Methods section as shown in Figure 3. Once REMY has completed, the 'Run' button updates to read 'Completed.'

### Data Reading

Beyond the standard codebase, necessary for the application execution, REMY also leverages spec2nii[29] and pymapVBVD (https://github.com/wtclarke/pymapvbvd) to read the various file formats. When reading Siemens data, REMY uses the 'pymapVBVD' function from spec2nii to read Twix and the 'multi_file_dicom' function from spec2nii to read dicom. For GE, the 'pfile' class and '_dump_struct' function are used to read pfiles. Philips spar/sdat is read using the read_spar function from spec2nii. REMY reads Bruker method files directly by parsing the text. Lastly, MRS-NIfTI files are read by parsing the JSON sidecar header file (.nii/.json).

### Header Fields

When reading each file, REMY attempts to identify as many relevant fields as possible from the data header. Field strength, vendor, vendor software, pulse sequence, number of data points, TE, TR, spectral width, and voxel size are among the parameters commonly available for each vendor. While each header file uses unique nomenclature, REMY identifies, translates, and inputs these parameters into the generic MRSinMRS format and table.

### Base Files

Included with REMY are a set of base files, each named MRSinMRS, that are used as a generic blueprint when generating the final outputs. The first of these files is the empty MRSinMRS table (.csv) that is read and populated throughout the execution process. Next, while LaTeX is not required to use REMY, all the necessary files to generate a LaTeX PDF are saved, including auxiliary document information (.aux), bibliography (.bbl), citations (.bib), control file (.bcf), and bibliography log (.blg). These files are each copied over to the export directory and renamed to reflect the name of the dataset that was inputted.

### Outputs

After completion, REMY will output a series of files to the export directory, each named after the input dataset. The first of which is a log file (.log) that includes detailed information about the runtime and notes whether any errors occurred. Next is the completed MRSinMRS table (.csv). If the users' system includes LaTeX, the necessary support files, LaTeX (.tex), and a text document (.pdf) will automatically be generated that now includes the information from the dataset.

It is increasingly recognized that the variability in data analysis methods among different vendor-proprietary pipelines and offline analysis tools cause substantial analytical variability, i.e., metabolite estimates are sensitive to the decisions made during pre-processing, modeling, and quantification.[26–28] The MRSinMRS table therefore contains sections requiring details of data analysis procedures and data quality, that for now, is not automatically populated by REMY because it is not linked to a processing pipeline/software. However, when integrated into existing software packages, this functionality would be easily implemented in the future.[3]

### Development Set

For the development of the REMY, an initial single voxel spectroscopy (SVS) dataset was used, consisting of 15 Siemens datasets including 8 twix (.dat), 3 DICOM (.ima) and 2 RDA (.rda) datasets, 8 Phillips .spar/.sdat datasets, 8 GE pfiles (.7) and 2 Bruker method files (PV 360 V1.1 and PV 360 V3.3). The initial development data came from 4 sites and was acquired at 3 field strengths (3T, 7T and 14T) with 6 different pulse sequences (STEAM, LASER, sLASER, PRESS, MEGA-PRESS and HERMES) ensuring a diverse development dataset covering the majority of SVS sequences (Supplementary Table 1).

After the application was developed, we performed a crash test using publicly available data from the Big GABA dataset (https://www.nitrc.org/projects/biggaba/). We randomly selected 15 PRESS datasets from 3 sites per vendor (i.e., 45 datasets per vendor) which contained GE (.7), Siemens Twix (.dat) and Philips (.spar/.sdat) files. We further expanded our test set with

spec2nii test data employing additional 4 GE (.7), 6 Philips (.spar/.sdat), 14 Siemens Twix, 2 Siemens DICOM (.ima), 3 RDA (.rda) and 2 Bruker (method) datasets to increase the diversity of sites and sequences included.

## 3. Results

Initial test data

The application was tested using an initial SVS test dataset (Siemens Twix (.dat) and Siemens DICOM (.ima), Phillips (.spar/.sdat), GE pfiles (.7), Bruker (.method), and MRS-NIfTI (.nii/.json) files) with the aim of reading, and translating all the available hardware and acquisition parameters required in the MRSinMRS table. For the hardware section of the table, REMY successfully read Field strength [T], Manufacturer name, Software version, and nuclei from all the file types, therefore populating 4 of 5 required hardware fields from the MRSinMRS table. RF coil and additional hardware information were not present in any of the data files and therefore could not be inputted.

For the “Acquisition” section of the table, pulse sequence name, nominal VOI size, TR, TE, total number of acquisitions or excitations per spectrum (NA), spectral width [Hz], and number of spectral points were successfully read from all file types populating an important part of the “Acquisition” fields of the MRSinMRS table. Necessary information on VOI anatomical location, WS method, transmit frequency offsets, scanner/software model and shimming method were not available in the majority of file types, but were read when possible. Therefore, the automatically generated methods section highlights all the necessary information that was not automatically populated in **bold**, indicating their importance and necessity for manual user input.

Test Data

List of parameters/fields that should have been filled: 1a, Field strength [T]; 1b, Manufacturer; 1c, Model; 1c, Software Version; 1d, Nuclei; 2a, Pulse Sequence; 2c, VOI Size; 2d, TR; 2d, TE; 2e, Total Number of Acquisitions (Total Excitations; NA); 2f, Spectral Width; and 2f, Number of Points. Supplementary Table 1 presents a comprehensive inventory of the datasets evaluated, along with the results of populating the necessary fields.

**Phillips .spar/.sdat -** In total, 21 Philips .spar/.sdat datasets were tested, ranging from software versions 3.2.1/.2.1 to 5.1.7/1.7. The application achieved a 100% success rate in importing the data and extracting the targeted parameters.

**Siemens Twix (.dat) -** 29 Siemens .dat datasets were tested, ranging from software version syngo MR B17 to E11. The application was 100% successful in reading in all 28 datasets and extracting the aimed parameters.

**Siemens DICOM (.ima) -** 2 Siemens .ima datasets were tested, ranging from MR B17 to E11. The application was 100% successful in reading in all datasets and extracting aimed parameters.

**Siemens RDA (.rda) -** 3 Siemens .ima datasets were tested with syngo MR B17 to XA31. The application was 100% successful in reading all datasets and extracting the aimed parameters.

**GE pfiles (.7) -** 19 GE pfiles (.7) were tested, ranging from software versions HD16 to MR24. The application was 100% successful in reading in all 19 datasets. It was 100% successful in extracting the aimed parameters, except for 5 datasets (software version 16) where it failed to extract the number of spectral points. It's important to emphasize that the application reads GE pfiles for version 7 onwards, as supported by spec2nii since the parameter notation is drastically different in files before that version.

**Bruker method files -** 2 Bruker datasets were tested. The application was 100% successful in extracting the aimed parameters for these datasets. The data types that Bruker exports include ser and fid files; however, they are always accompanied by a method text file that contains all the necessary information for the MRSinMRS table. The nomenclature in the method files remains unchanged across versions (here tested PV6.0.1, PV 360 V1.1, PV 360 V3.3); therefore we chose to read this file type for our application.

**MRS-NIfTI .nii/.json -** Finally, one datasets from each available data type (6 datasets) was converted to the MRS-NIfTI data format using spec2nii (version 0.8.4) which were further used to test the MRS-NIfTI .nii/.json reader. The application was 100% successful in extracting the aimed parameters. It is important to note that the number of fields that can be populated from the MRS-NIfTI header is fewer than the original raw data file.

## 4. Discussion

REMY is a robust and convenient tool that enables researchers and clinicians to report essential MR hardware and acquisition parameters for MRS experiments. By automatically populating a standard table suggested by a consensus paper, sourced from a single MRS data file, the process facilitates straightforward study replication and streamlined method evaluation. Furthermore, the tool generates a methods section, simplifying the reporting process for researchers, which can help ensure the validity of the study setup and the interpretability of the results. The alternative requires a complete manual search for those parameters, and exporting and populating the table justified in the consensus paper.[15] While our REMY tool populates the hardware and acquisition sections of the MRSinMRS table from most data formats currently used (Figure 1), it cannot populate the other sections, and it is imperative for researchers to carefully fill those in manually after completing the analysis. This process can and should be automated as well; for example, the end-to-end analysis pipeline Osprey populates the data analysis methods and data quality sections in the MRSinMRS table with information generated by its built-in linear-combination modeling and quantification modules.[24] Leveraging the open-source and adaptable nature of REMY, we anticipate that other MRS analysis software developers may seamlessly integrate it into their pipelines. With straightforward modifications in the source code, the complete table generation process can be automated, further enhancing the ease and efficiency of MRS reporting.

While many have cited the MRSinMRS consensus paper, indicating the willingness of the MRS community to provide detailed methodological approaches, often the table is not included in neither the manuscript nor supplementary materials. A notable challenge in completing the table is the variation in parameter nomenclature among different scanner vendors, which can complicate the task of locating and reporting parameters. Additionally, given the inherent variations across vendors and software versions, we have established a detailed, frequently updated table on GitHub with software versions and datasets tested and implemented throughout REMY development. Other parameters like dynamics, transients, averages, and blocks also can be misconstrued as the same. In response, our application is equipped with a backend code that identifies the vendor and automates the population of parameters, mitigating the potential for human error. For instance, the total number of excitations or acquisitions per spectrum (2e) indicates how many single scans were recorded for the averaged spectrum. The subsequent fields 2e(i) - number of averaged spectra per time point (number of excitations per time point), 2e(ii) - averaging method and 2e(iii) total number of spectra, are required only for kinetic studies. By populating the table automatically, the confusion between terms has been successfully solved by REMY.

We have addressed the technical challenges that could discourage researchers from using REMY by developing a user-friendly, standalone application that does not require advanced coding skills. While we are confident in its ease of use, further testing with a wider range of users could help us refine and improve its accessibility. Note that for all vendors, the application successfully reads pulse sequence names; however, the sequences are often user modified and renamed for different studies. As such, the name that is given to the sequence and subsequently read by the application might not contain its original/true name and needs to be corrected by the user.

The application successfully translated the test data acquired with various MRS sequences to the output table and methods section as shown in Figure 5. Parameters that are used as a metric of success are field strength [T], manufacturer, model, software version, nuclei, pulse sequence, VOI size, TR, TE, NA, spectral width, and number of points. In cases where certain parameters are present in one data file format but absent in another, and consequently not filled in by REMY for the formats lacking this information, we interpret this as a failure or limitation in accurately reading and reporting parameters which has affected success rate measurement.

| MRSinMRS checklist | |
|---|---|
| Site (name or number) | . · |
| 1. Hardware | |
| a. Field strength [T] | 2.89 T · |
| b. Manufacturer | Siemens |
| c. Model (software version if available) | MAGNETOM Prisma |
| d. RF coils: nuclei (transmit/receive), number of channels, type, body part | 64 ch, 1H |
| e. Additional hardware | |
| 2. Acquisition | |
| a. Pulse sequence | PRESS |
| b. Volume of interest (VOI) locations | ACC |
| c. Nominal VOI size [cm3, mm3] | 40x25x27mm3 |
| d. Repetition time ($T_R$), echo time ($T_E$) [ms, s] | 30/2000 |
| e. Total number of excitations or acquisitions per spectrum | 64 |
| In time series for kinetic studies | |
| i. Number of averaged spectra (NA) per time point | |
| ii. Averaging method (eg block-wise or moving average) | |
| iii. Total number of spectra (acquired/in time series) | |
| f. Additional sequence parameters (spectral width in Hz, number of spectral points, frequency offsets) | 2000Hz, 2048 data points |
| If STEAM:, mixing time ($T_M$) | |
| If MRSI: 2D or 3D, FOV in all directions, matrix size, acceleration factors, sampling method | |
| g. Water suppression method | |
| h. Shimming method, reference peak, and thresholds for "acceptance of shim" chosen | BRAIN |
| i. Triggering or motion correction method (respiratory, peripheral, cardiac triggering, incl. device used and delays) | |

Methods provided by the automatic table tool based on MRSinMRS reporting standards

Antonia Susnjar, Antonia Kaiser, Gianna Nossa, Dunja Simicic, Aaron Gudmundson

January 24, 2024

1 Methods

1.1 $^1$H-MRS Spectroscopy

1.1.1 Acquisition

All scanning procedures were conducted using a Siemens MAGNETOM Prisma scanner equipped with a 64ch coil operating at 2.89 T. Before spectroscopy data acquisition, a **sequence for anatomical image** anatomical image **(info)** was obtained for accurate MRS voxel **placement and subsequent** segmentation. MRS data were acquired **utilizing** the PRESS sequence in combination with **WS method** water suppression method. The following acquisition parameters were used: repetition time/echo time (TR/TE) of 2000 ms/30 ms, and 64 transients recorded per acquisition of **total number of acquisitions** and a spectral width of 2000 kHz. The volume of interest (VOI = 40 x 25 x 27) was placed in ACC. The static magnetic field homogeneity was adjusted using BRAIN shimming method, the threshold for the acceptance of shim was set at **thresholds** using **reference peak** as the reference peak. For eddy current correction and quantification purposes, **number of averages** transients of water-unsuppressed data were also acquired, using identical acquisition parameters.

Note: !!! The positions in the text which could not be automatically filled using the table are highlighted in **BOLD**. Please fill in this information manually. !!! If you used multiple VOI or various VOI locations with the same protocol, please change the text accordingly to include VOI specific information (both location and threshold for shim).

- IF Diffusion: !!! If you use a diffusion weighted sequence, please include additional information: diffusion time (), b-values, number of transients per b-value and the diffusion direction.

Figure 5. Exported output from the REMY standalone application using a Siemens .twix MRS dataset. The first two sections of the table regarding data acquisition and hardware are populated, while data processing and quality need to be manually inputted by the researchers.The right side of Figure 5 shows the automatically generated methods section that researchers can directly copy into their manuscript. This section includes parameters extracted from the MRS file header, simplifying the process of incorporating standardized reporting into their work.

One major constraint of this project is its limited use for single voxel MRS, which inherently minimizes applicability across a broader spectrum, including multi-voxel MRS studies. However, all code is openly available, to encourage the community to contribute to REMY and extend its usability to all possible MRS cases.

The rapid increase in development of available MRS processing software presents an opportunity for considerable advancement within the MRS field. As mentioned before, a drawback resides in the need for conclusive reporting of methods. We are actively strategizing to extend our outreach to software developers working on the development of MRS processing tools. The essence of our plan involves integrating our open-source application into existing tools, creating a partnership that simplifies reporting methods for the MRS scientists and eliminates an aspect of what can be perceived as “black box” processing. Although many MRS processing tools have taken steps to ensure transparency and logical checkpoints throughout their processing pipelines, the inclusion of MRSinMRS outputs and the completion of the third section in the table, pertaining to data analysis methods and outcomes, will further enhance the credibility of published research.

Furthermore, we provide a standardized and automatically filled methods section with adequate references. This not only establishes a standardized template, but also provides a concrete methodological framework that can be directly incorporated into publications. Given the increasing number of published MRS studies, we recognized a challenge within the field and provided a simple solution through our work. REMY as an application is a start to provide a valuable, easy-to-use, expandable resource for novice and expert MRS researchers.

**Future Directions**

Despite its current limitation to single voxel multinuclear MRS studies, the open-source nature of REMY invites contributions that could extend its applicability to other MRS applications. Moreover, the ability of REMY to be integrated into existing MRS processing tools promises further simplification of methodological reporting, increasing the transparency and credibility of MRS research.

Lastly, the implementation of REMY could be refined, and the source code could be improved in accordance with community recommendations. For example, the future roadmap for REMY includes a demographics section that will be generated when researchers import all datasets of one study. The demographics section would provide the number of participants and, where available, gender, mean age, etc. This would be especially helpful for meta/mega-analyses. With the input of all datasets, REMY would highlight fields where the parameters, accidentally or deliberately, diverge across scans. In addition, after extraction is complete, the user would be presented with an editable list of the extracted parameters so that they can insert parameters that are typically not in many headers (i.e., water suppression method) before the methods section is generated.

## 5. Conclusion

We developed the REMY application to boost the adoption of MRSinMRS consensus recommendations as a solution for insufficient MRS research reporting. By automating the population of the MRSinMRS table with essential study parameters from a single dataset, REMY facilitates the replication of studies and evaluation of methodologies. Challenges such as the variation in parameter nomenclature across different scanner vendors and the technical hurdles associated with manual table completion have been effectively resolved. Additionally, by providing an automatically filled, standardized methods section with appropriate references, REMY sets a template that can be directly incorporated into publications, addressing the pressing need for standardized reporting in the growing body of MRS literature.

**Acknowledgments:**

We would like to thank the following research sites for their contribution of deidentified data:, Purdue University, Spec2nii[29], BIG GABA[30],École Polytechnique Fédérale de Lausanne (EPFL), CIBM Center for Biomedical Imaging, and Amsterdam University Medical Center (AUMC).

Our gratitude extends to the organizers and hosts of the inaugural MRS Hackathon 2023, who afforded us both space and the chance to interact with a diverse group of MR spectroscopists. This collaborative environment enabled us to conceive and actualize the REMY application.

Their support was instrumental in converting our concept into a concrete solution, addressing the gap of reporting standards in the MRS field, which emerged during discussions at the MRS Workshop 2022 in Lausanne, Switzerland.
We would like to acknowledge the ResearchHub grant for funding our project that will support completing implementations discussed in the future directions of this paper. GO acknowledges funding from NIH grants R00 AG062230, R21 EB033516, and P41 EB031771.

**CRediT Author Contributions:**
**Antonia Šušnjar:** Conceptualization, Data Curation, Funding Acquisition, Investigation, Methodology, Project Administration, Supervision, Visualization, Writing – Original Draft Preparation. **Antonia Kaiser:** Conceptualization, Data Curation, Methodology, Investigation, Software, Supervision, Writing – Original Draft Preparation. **Dunja Simicic:** Data Curation, Formal Analysis, Investigation, Software, Validation, Visualization, Writing – Original Draft Preparation. **Gianna Nossa:** Investigation, Writing – Original Draft Preparation, Writing – Review & Editing. **Alexander Lin:** Project Administration, Supervision, Writing – Review & Editing. **Georg Oeltzschner:** Project Administration, Supervision, Writing – Review & Editing. **Aaron Gudmundson:** Conceptualization, Methodology, Project Administration, Software, Supervision, Validation, Visualization, Writing – Review & Editing.